\documentclass[twocolumn,english,aps,superscriptaddress,prl,floatfix,longbibliography]{revtex4-2}
\usepackage{scalefnt}
\usepackage[utf8]{inputenc}
\usepackage{graphicx}
\usepackage{dcolumn}
\usepackage{braket}
\usepackage{amsthm}
\usepackage{bm}
\usepackage{float}

\usepackage{multirow}
\usepackage{wrapfig}
\usepackage{amsmath}
\usepackage[shortlabels]{enumitem}
\usepackage{amsfonts}
\usepackage{physics}
\usepackage{braket}
\usepackage[makeroom]{cancel}
\usepackage{mathrsfs}
\usepackage{ gensymb }
\usepackage{dsfont}
\usepackage{amssymb}
\usepackage[title]{appendix}

\usepackage[dvipsnames]{xcolor}
\usepackage[mathscr]{euscript}
\usepackage{yfonts}
\usepackage[export]{adjustbox}
\usepackage{pagecolor}
\usepackage{tcolorbox}
\usepackage{diagbox}
\usepackage{booktabs}
\usepackage{color, colortbl}

\usepackage[unicode=true,pdfusetitle,
bookmarks=true,bookmarksnumbered=false,bookmarksopen=false,
breaklinks=true,pdfborder={0 0 0},pdfborderstyle={},backref=false,colorlinks=true]
{hyperref}
\hypersetup{
colorlinks=true,
linkcolor=blue,
filecolor=magenta,      
urlcolor=blue,
citecolor=blue
}

\newcommand{\pa}[1]{\left(#1\right)}
\newcommand{\co}[1]{\left[ #1\right] }

\newcommand{\urz}{\textbf{u}_0(\textbf{r})}

\newcommand{\uv}[1]{\textbf{u}_{#1}}

\newcommand{\q}{\textbf{q}}

\begin{document}

\title{Raman scattering from moir\'e phonons}
\author{Vitor Dantas}
\affiliation{School of Physics and Astronomy, University of Minnesota, Minneapolis, MN 55455, USA}
\author{Héctor Ochoa}
\affiliation{Department of Physics, Columbia University, New York, NY 10027, USA}
\author{Rafael M. Fernandes}
\affiliation{Department of Physics, The Grainger College of Engineering,
	University of Illinois Urbana-Champaign, Urbana, IL 61801, USA}
\affiliation{Anthony J. Leggett Institute for Condensed Matter Theory, The Grainger College of Engineering,
	University of Illinois Urbana-Champaign, Urbana, IL 61801, USA}
\author{Natalia B. Perkins}
\affiliation{School of Physics and Astronomy, University of Minnesota, Minneapolis, MN 55455, USA}

\begin{abstract}
We develop a theoretical framework for probing moir\'e phonon modes using Raman spectroscopy, and illustrate it with the example of twisted bilayer graphene (TBG). These moir\'e phonons arise from interlayer sliding motion in twisted 2D materials and correspond to fluctuations of the stacking order in reconstructed moir\'e superlattices. These include both acoustic-like phason modes and a new set of low-energy optical modes originating from the zone-folding of monolayer graphene's acoustic modes, which are accessible via Raman spectroscopy. We show that the Raman response of TBG exhibits a series of low-frequency peaks that clearly distinguish it from that of decoupled layers. We further examine the role of anharmonic interactions in shaping the phonon linewidths and demonstrate the strong dependence of the Raman spectra on both the twist angle and the polarization of the incident light. Our findings establish Raman spectroscopy as a powerful tool for exploring moir\'e phonons in a broad class of twisted van der Waals systems.

\end{abstract}
\date{\today}

\maketitle

%
%
{\color{blue}\textit{Introduction}}--
The discovery of moir\'e superlattices in twisted 2D materials has led to a diverse range of emergent phenomena, including correlated insulating states, topological phases, superconductivity, and magnetism \cite{Cao2018Correlated,Cao2018Unconventional,Yankowitz2019,Sharpe2019,Jin2019,Tang2020,Zeng2023,Xia2025,Guo2025,Park2023}.
While much of the focus has been on electronic properties, recent studies have highlighted the important role of lattice reconstruction \cite{nam2017lattice,carr2018relaxation,GuineaWalet2019,Wang2024,KangVafek2025,DeBeule2025} and of moir\'e phonons \cite{koshino2019moire,ochoa2019moire,Bernevig2019,DasSarma2019,phasons_TMD,xiao2021chiral,samajdar2022moire,lu2022low,GaoKhalaf2022,Birkbeck2024Phason,Xiao2024,Ochoa_linear2023,Girotto2023,Guinea2023,Parameswaran2024,Maity2023,Huang2025,Ramos-Alonso2025} in shaping the low-energy physics of these systems. 
The moir\'e phonon spectrum features low-energy interlayer shear modes corresponding to the relative sliding between layers, 
\cite{koshino2019moire,ochoa2019moire,ochoa2022degradation}. Additional contributions appear at higher energies, coming from symmetric and antisymmetric out-of-plane breathing modes, as well as intralayer optical phonons that hybridize and fold due to the moiré periodicity, forming mini-phonon bands. These emergent modes are shaped by the moir\'e superlattice, which itself results from the competition between the interlayer adhesion potential and the intralayer elastic energy, and are collectively referred to as moir\'e phonons.


In this Letter, we develop a theoretical framework for the Raman response of moir\'e  phonons and apply it to twisted bilayer graphene (TBG). Our focus is on the low-energy region of the moir\'e phonon spectrum, which arises from the folding of the non-Raman active acoustic modes of single-layer graphene. Thus, their Raman response should encode unique information about the moir\'e superlattice, e.g. the twist angle and the strength of the adhesion potential. In particular, we are interested in how these low-energy moir\'e  phonons couple to light, with a focus not only on understanding their signatures in Raman spectroscopy but also on exploring how this coupling is influenced by the underlying superlattice structure and twist angle. While there have been some efforts in this direction, both in twisted bilayer graphene (TBG) \cite{Carozo2011,Kim2012,Sato2012,Carozo2013,E2013,Campos2013,Schapers2022,Gadelha2021,Barbosa2022,Pandey2024} and in transition metal dichalcogenides (TMDs) \cite{Lin2021,Parzefall_2021,Wu2023}, a systematic experimental investigation and   a comprehensive theoretical framework remain lacking.

In what follows,
we show that the Raman response of TBG features a series of low-frequency peaks arising from zone-folded shear modes, with frequencies, intensities, and polarization dependence that encode detailed information about the twist angle and moir\'e symmetry. These include strong intensity variations among modes of the same symmetry, reflecting differences in their phonon wavefunctions. By combining symmetry analysis with a continuum elasticity model, we derive explicit expressions for the Raman tensor and compute the full Raman spectra, thus providing important guidance for future experimental observations.

\begin{figure*}
	\centering
	\includegraphics[width=1\linewidth]{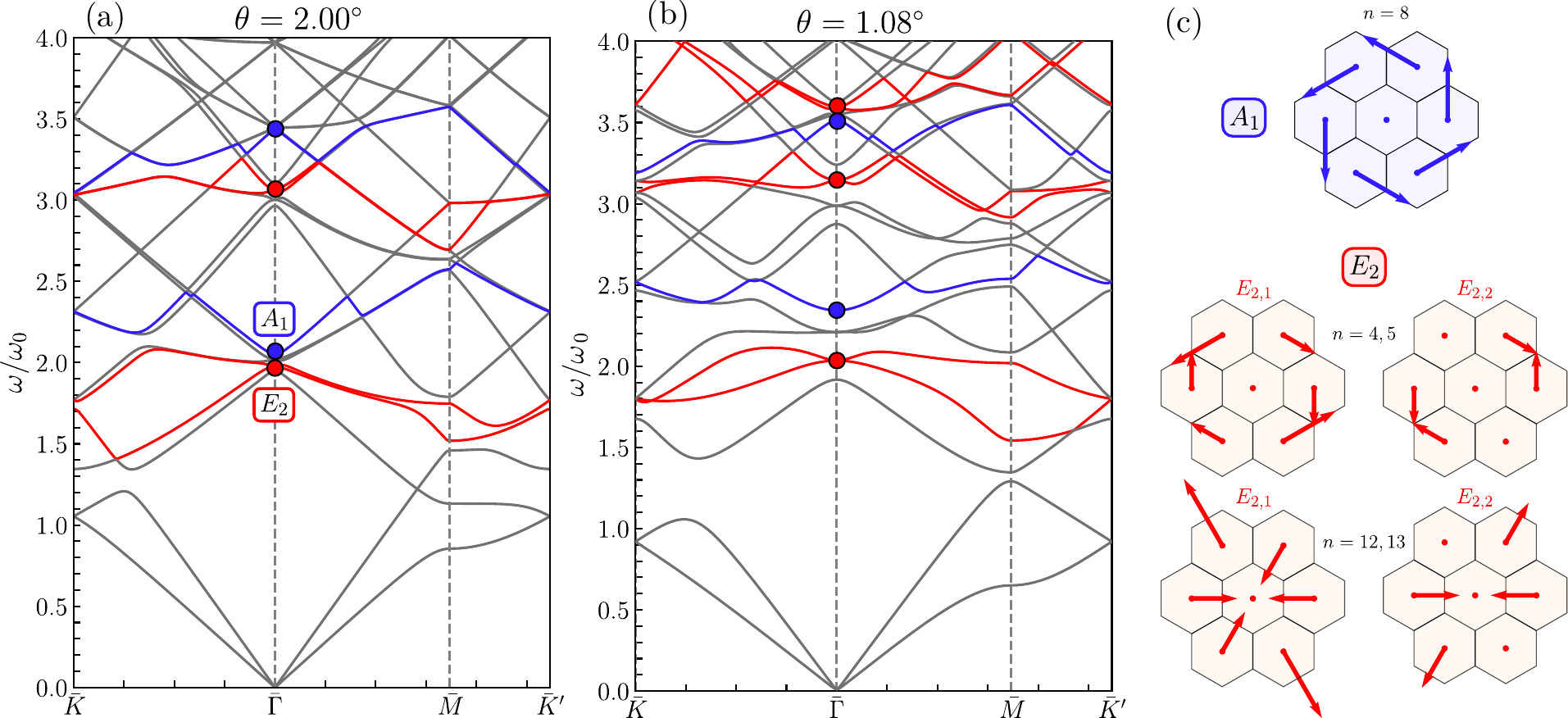}
	\caption{ Moir\'e phonon dispersion along the high-symmetry path in the mBZ for $\theta = 2.0^\circ$ (a) and $\theta = 1.08^\circ$ (b). The frequencies are scaled by the parameter $\omega_0 = (2\pi/L_M)\sqrt{\lambda/\rho}$, which takes values of $4.86$ meV and $2.63$ meV, respectively.  The in-plane Raman active modes at  $\textbf{q}=0$ are indicated by filled circles, blue for $A_1$ and red for $E_2$ doublets. In (c) we show the layer-shear displacements $\delta \textbf{u}_{\textbf{G}}$ corresponding to the lowest $A_1$ and $E_2$ modes in moir\'e reciprocal space with vectors  $\textbf{G}$ in the moir\'e first star. We denote the partners of each $E_2$ mode as $E_{2,1}$ and $E_{2,2}$. The branches $n$ are indicated for each mode, corresponding to the highlighted bands in (a) and (b) for the respective colors. The gray curves correspond to phonon branches that are not Raman-active in the chosen backscattering geometry.}
	\label{fig:moiredispersion1}
\end{figure*}

%
%
{\color{blue}\textit{Lattice relaxation and moiré phonons}} -- 
The first step in computing the moir\'e phonon spectrum is to determine the structural reconstruction of the TBG lattice.  
Since our primary interest is in the contribution of the folded sliding modes, we adopt a continuum approach based on standard elasticity theory, as outlined in Refs. 
\cite{ochoa2019moire, koshino2019moire,ochoa2022degradation}, and provide additional details in the Supplemental Material (SM) \cite{suppl}.

 The system's free energy is written 
 as a functional of the lateral displacement between the two rotated graphene layers, defined as $\textbf{u}(\textbf{r}) = \textbf{u}^{(2)}(\textbf{r}) - \textbf{u}^{(1)}(\textbf{r})$,  where the superscript denotes the layer index. This displacement field determines the local stacking configuration, which is assumed to vary smoothly in space. The total free energy consists of two contributions, $\mathcal{F}[\textbf{u}] = \mathcal{F}_{\text{el}}[\textbf{u}] + \mathcal{F}_{\text{ad}}[\textbf{u}]$, where the first term is the elastic energy associated with in-plane distortions, and the second term is the adhesion energy due to interlayer coupling. Explicitly, the elastic term is given by
\begin{align}
	\mathcal{F}_{\text{el}}[\textbf{u}] = \int d^2r   \co{ \frac{\lambda}{2}(\nabla \cdot \textbf{u})^2 + \frac{\mu}{4}\pa{ \partial_iu_j + \partial_ju_i }^2   },
\end{align}
where $\lambda \approx 3.25 \text{eV/\AA}^2$ and $\mu \approx 9.57 \text{eV/\AA}^2$ are the Lam\'e coefficients of graphene \cite{zakharchenko2009finite,lee2008measurement}. The  adhesion potential term, $\mathcal{F}_{\text{ad}}$, accounts for the variation in energy between different stacking configurations due to interlayer coupling. By symmetry, 
the leading-order Fourier expansion of
 $\mathcal{F}_{\text{ad}}[\textbf{u}]$ takes the form:
\begin{align}
	\mathcal{F}_{\text{ad}}[\textbf{u}] = V_0 \int d^2r \sum_{j=1}^3 \cos \pa{ \textbf{G}^M_j \cdot \textbf{r} + \textbf{u}(\textbf{r})\cdot\textbf{b}_j},
\end{align}
where $\textbf{G}^M_j$ are the moir\'e reciprocal lattice vectors and $\textbf{b}_j$ are the corresponding graphene reciprocal lattice vectors. This term penalizes local AA stacking and favors AB stacking, reflected by their energy difference $V_0$. 
Following Refs.\cite{carr2018relaxation,ochoa2019moire}, we set the parameter $V_0\approx 90 \text{meV/nm}^2$
 as the strength of the adhesion potential.  
 Since the interlayer potential depends only on the relative displacement field $\textbf{u}(\textbf{r})$, center-of-mass motions can be safely neglected. The relaxed structure is then obtained by minimizing the total free energy. 
  We numerically solve this problem in a self-consistent manner (See SM \cite{suppl}), and denote the resulting displacement field by $\textbf{u}_0(\textbf{r})$.

  After obtaining the optimal stacking configuration for a given twist angle $\theta$, we introduce dynamical fluctuations around the relaxed structure via $\delta \textbf{u}(\textbf{r},t) = \textbf{u}(\textbf{r},t) - \textbf{u}_0(\textbf{r})$. This fluctuation field can be expanded in Fourier components as
   $\delta \textbf{u}(\textbf{r},t) = \sum_{\textbf{q} \in \text{mBZ}}\sum_\textbf{G} e^{i(\textbf{q}+\textbf{G})\cdot\textbf{r}}\delta \textbf{u}_{\textbf{q}+\textbf{G}}(t)$,
   where mBZ denotes the moir\'e Brillouin zone. 
    The free-energy cost of these stacking fluctuations can be computed in the harmonic approximation, giving us the dynamical matrix $\mathbb{D}_\textbf{q}$, where $\textbf{q} \in \text{mBZ}$. Diagonalizing $\mathbb{D}_{\textbf{q}}$ (See SM \cite{suppl}) gives the moiré phonon spectrum $\omega_{n,\textbf{q}}$, consisting of two acoustic-like phason modes  and $2N_G - 2$ optical branches, where $N_G$ is the number of $\mathbf{G}$ vectors considered in the calculations. The resulting phonon dispersions for three representative twist angles are shown in Fig.~\ref{fig:moiredispersion1}.

The gapless modes correspond to phasons \cite{ochoa2019moire}, which become gapped and overdamped in the presence of impurities \cite{ochoa2022degradation,Ochoa_linear2023}. Phasons will not be studied here, as they are not observed from Raman scattering. The
normalized eigenvectors of $\mathbb{D}_\textbf{q}$, denoted as $\textbf{C}_{n,\textbf{q}}(\textbf{G})$ (See SM \cite{suppl}), provide a basis for a normal mode expansion \cite{koshino2020effective}:
\begin{align}\label{delta_u}
	\delta \textbf{u}(\textbf{r},t) = \sum_{\textbf{q}\in \text{mBZ}}\sum_{\textbf{G}, n} \,  e^{i\pa{\textbf{G} + \textbf{q}}\cdot \textbf{r}} \textbf{C}_{n,\textbf{q}}(\textbf{G})\, \hat{Q}_{n,\textbf{q}}(t),
\end{align}
where $n$ are the optical phonon branches. Here we introduced the second-quantized normal modes coordinates $Q_{n,\textbf{q}} = \sqrt{\frac{\hbar}{2A \rho \omega_{n,\textbf{q}}}}\, \pa{ a_{n,-\textbf{q}}^\dagger + a_{n,\textbf{q}} }$, with $\rho = 7.6 \times 10^{-7}$ kg/m$^2$ as the mass density of graphene and $A$ is the moiré unit cell area.

%
%

{\color{blue}\textit{Microscopic theory of Raman scattering}}-- 
Raman spectroscopy is based on the inelastic scattering of light, where incident photons with momentum $\textbf{k}=0$ interact with collective excitations of the system, such as phonons, plasmons, or magnons, leading to a frequency shift in the scattered light \cite{poulet1970spectres,hayes1978scattering}. Here we are interested in the scattering from  moiré phonons only.  Without loss of generality, we restrict our analysis to Stokes processes, so the frequency shift is positive. The incident electric field $\textbf{E}_\text{in}$ induces a time-dependent polarization in the medium, $P^\mu(t) = \chi^{\mu\nu}(t) E^\nu_\text{in}$, where $\chi^{\mu\nu}(t)$ is the polarizability tensor.
Fluctuations in the stacking texture modulate this tensor, thereby enabling inelastic scattering. 

In analogy to the adhesion potential, we express the stacking‑dependent polarizability tensor as a series of moiré harmonics (See SM \cite{suppl}).
\begin{align}\label{polarizability_moire}
	\chi^{\mu\nu}[\textbf{u}(\textbf{r},t)]\sim \sum_{s,j}[\hat{\mathbf{G}}^s_j]^{\mu} [\hat{\mathbf{G}}^s_j]^{\nu}\,\cos  \pa{\textbf{G}^s_j \cdot \textbf{r} + \textbf{u}\cdot\textbf{b}_j^s },
\end{align}
where $s$ indicates the stars of the moiré and graphene reciprocal spaces and the unit vectors ensure the symmetry constraints on $\chi^{\mu\nu}(t)$. We consider the linear response regime, so $\chi^{\mu\nu}(t) = \chi^{\mu\nu}_{\text{Rayl.}} +  \chi^{\mu\nu}_{\text{Ram.}}(t)$. The zero-th order term is responsible for elastic (Rayleigh) scattering, while the second is the contribution to the Raman scattering. In terms of the moir\'e phonon field, this contribution is $	\chi^{\mu\nu}_{\text{Ram.}}(t) = \int d^2\textbf{r}\,\,\widetilde{R}^{\mu\nu}_{\lambda}(\textbf{r})\,\delta u_\lambda(\textbf{r},t),$ where  $\widetilde{R}^{\mu\nu}_{\lambda}(\textbf{r}) = \pa{\partial{\chi^{\mu\nu}} /\partial {u_\lambda(\textbf{r})}}_0$  is a rank-3 Raman coupling tensor evaluated at equilibrium, which accounts for the modulations of $\chi^{\mu\nu}$ in the stacking order fluctuations.

It is possible to re-express this response in terms of the normal modes, $\chi^{\mu\nu}_{\text{Ram}} = \sum_n R_{n}^{\mu\nu}Q_n(t)$,  where we define the moir\'e phonon Raman tensor $R^{\mu\nu}_n \equiv \pa{\partial{\chi^{\mu\nu}} /\partial {Q_n}}_0$. This is in complete analogy to the Raman response in crystalline solids, where the material’s polarizability couples linearly to the normal mode coordinates of the phonons, and the independent components of $R^{\mu\nu}_n$ are determined from a group-theoretical analysis \cite{hayes1978scattering,poulet1970spectres}. 

From the expression in 
Eq.\eqref{polarizability_moire} we can obtain an explicit form of the Raman tensor components,
\begin{align}\label{eq:Raman_tensor}
	R^{\mu\nu}_n = \sum_{s,i}\sum_{\textbf{G}} \zeta_s \,\, C^\lambda_n(\textbf{G})\,  [\hat{\textbf{b}}^{s}_i]_{\lambda}[\hat{\mathbf{G}}^s_i]^{\mu} [\hat{\mathbf{G}}^s_i]^{\nu} f_i^s(\textbf{G}),
\end{align}
where we sum over the stars up to $N_\text{star}$, $i=1$ to 3 and the function $ f_i^s(\textbf{G})$ is defined in the SM \cite{suppl}. Here the constants $\zeta_s$ are dimensionless amplitudes that weight the $s$-th moiré harmonic of $\chi^{\mu\nu}$. We expect modes with the moiré length scale to dominate the response, so it is reasonable to assume that $\zeta_s$ decay as $\zeta_s \sim |\textbf{G}^M_j|/\abs{\textbf{G}^s_j}$. Consequently, the first star gives the dominant contribution to the Raman intensity, while higher stars give only small corrections.

To compute the scattered intensity, we define the \textit{Raman vertex operator} as $\mathcal{R}(t) = \sum_{\mu\nu}E_{\text{in}}^\mu E_{\text{out}}^\nu \,  \chi^{\mu\nu}_{\text{Ram.}}(t)$, where 
$E_{\text{in}}^\mu$ and $E_{\text{out}}^\nu$ are the components of the incident and scattered electric fields, respectively.  The Raman intensity is then given by the time-ordered correlation function of the vertex operator:  $\mathcal{I}(\omega) = \int dt e^{i\omega t}  \expval{T\, \mathcal{R}(t)\, \mathcal{R}(0)}$.
The explicit form of the Stokes intensity can be obtained through the fluctuation-dissipation theorem:
\begin{align}\label{eq:Raman_intensity}
	\mathcal{I}(\omega) \propto  \sum_n \abs{ \sum_{\mu\nu} E_{\text{in}}^\mu \,  R^{\mu\nu}_{n} \, E_{\text{out}}^\nu }^2 \Im D^R_n(\omega, \textbf{q}=0), 
\end{align}
where $D^R_n(\omega, \textbf{q}=0)$ is the retarded Green's function of moiré phonons in the branch $n$, and we omit the temperature dependence for simplicity.



\begin{figure*}
	\centering
 	\includegraphics[width=1\linewidth]{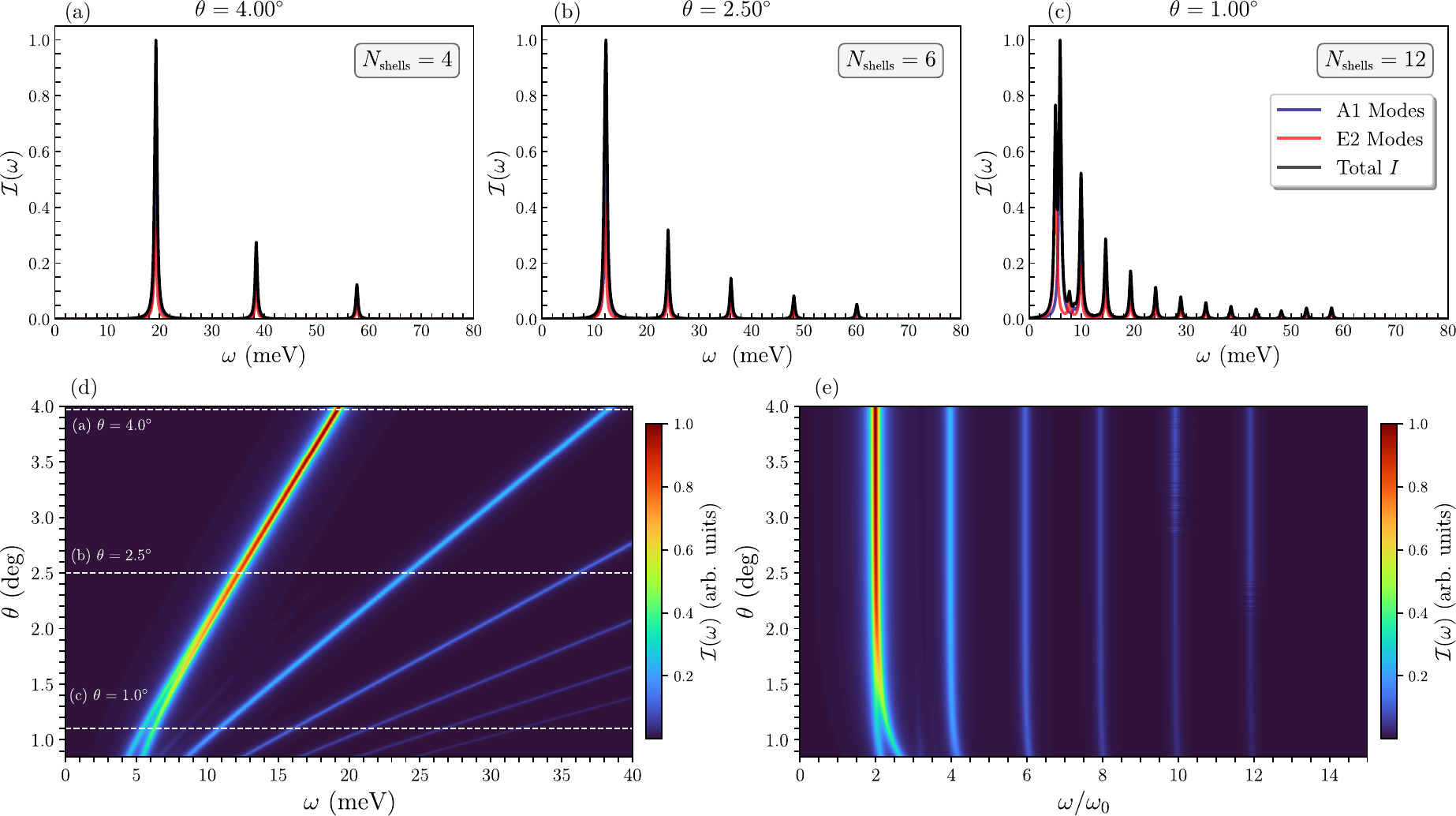}
	\caption{(a)-(c) Raman intensity in arbitrary units. Here we choose the $\bar{z}(xx)z$ geometry, where both the $A_1$ and $E_2$ modes contribute to the response. Visually, the response from $A_1$ and $E_2$ almost overlaps because the gap between these modes is very small. For each angle we choose a number of $\textbf{G}$ shells such that the Raman spectrum is computed up to $ \omega_n \sim 80$ meV, as indicated on the top right of each plot. The broadening for all angles is $\gamma = 0.1$ meV. In (d) we show how the Raman intensity evolves with the angle in the range $ \theta \in(0.85^\circ,4^\circ)$ in the small frequency limit ($\omega \lesssim $ 40 meV). The white lines correspond to the cuts shown in (a)-(c). In (e) we rescale the frequencies by $\omega_0 = (2\pi/L_M)\sqrt{\lambda/\rho}$, where the approximate linear relation between $\theta$ and $\omega$ is evidenced. For small angles the relaxation is stronger, so the lowest Raman peaks acquire a non-linear dependence on $\theta$.}
	\label{fig:ramanintensity}
\end{figure*}

%
%

{\color{blue}\textit{Symmetry classification of Raman-active  modes}}--
In crystalline systems, the form of the Raman tensor is dictated by lattice symmetries, with selection rules determined by the point group at $\mathbf{q} = 0$. In moiré systems, the Raman tensor $R^{\mu\nu}_n$ must similarly respect the symmetries of the emergent moiré pattern, which for relaxed TBG corresponds to the chiral point group $D_6$. Due to the lack of inversion symmetry in $D_6$, certain infrared-active modes can also be Raman-active \cite{aroyo2006bilbao,poulet1970spectres}. In this work, we focus on backscattering geometry, where among the in-plane vibrational modes $A_1$, $E_1$, and $E_2$, only the $A_1$ and $E_2$ modes contribute to the Raman response.


To identify which modes at $\textbf{q}=0$ are Raman-active, we thus must classify the phonon eigenvectors according to the irreducible representations of the $D_6$ point group. 
Specifically, for each eigenvector $\textbf{C}_{n,\textbf{q}=0}(\textbf{G})$, we construct projection matrices onto the irreps of $D_6$. 
 This is done by building a higher-dimensional (reducible) representation of $D_6$ that includes all possible permutations of $\textbf{G}$ points under the symmetry operations, together with the transformations of the displacement vectors $\delta \textbf{u}_{\textbf{G}} = (\delta u^x_\textbf{G},\delta u^y_\textbf{G})$ \cite{dresselhaus2007group}. The full representation is given by the tensor product $
\widetilde{\Gamma} = \Gamma_{\text{perm}} \otimes \Gamma_{\text{vec}}$ with symmetry operations $g \in D_6$ represented as $D^{\widetilde{\Gamma}}(g) = 	D^{\Gamma_{\text{perm}}}(g) \otimes 	D^{\Gamma_{\text{vec}}}(g)$. For a given number of stars of $\textbf{G}$, the decomposition gives the expected number of Raman-active modes as
 $\widetilde{\Gamma}_\text{Raman} = N_{\text{star}}(A_1\oplus 2E_2)$. 

The projection operator onto each irrep 
 $\Gamma_j$ of $D_6$ is defined as
\begin{align}\label{eq:projection_charac}
	\Pi^{(\Gamma_j)} = \frac{d_j}{12}\sum_{g\in D_6}\co{\chi^{(j)}(g)}^*D^{\widetilde{\Gamma}}(g),
\end{align}
where $d_j$ is the dimension of $\Gamma_j$ and $\chi^{(j)}(g)$ are  the corresponding characters. The phonon eigenmode is represented as a $2N_G$-dimensional array, $\boldsymbol{\xi}_n^T = \co{ \textbf{C}_n(\textbf{G}_0) \, , \,  \textbf{C}_n(\textbf{G}_1) \, , \,  \cdots, \textbf{C}_n(\textbf{G}_{N_G}) }$, which encodes the displacement directions for mode $n$. A mode transforms as $\Gamma_j$ if $\Pi^{(\Gamma_j)} \boldsymbol{\xi}_n \neq 0$.
The resulting set of in-plane Raman-active modes for the first shell of $\textbf{G}$ vectors is shown in Fig. \ref{fig:moiredispersion1}(c). Notably, the symmetric mode $A_1$ is purely transverse, in contrast to the typical breathing-like vibration often associated with $A_1$ modes in crystals. This comes from the fact that $\delta \textbf{u}_\textbf{G}$ is a relative displacement between the layers, so it must be odd under the two-fold rotations $\mathcal{C}_{2x}$ and $\mathcal{C}_{2y}$.

%
%


%
%

{\color{blue}\textit{Results for the Raman intensity in TBG}}--  Here we present our results for the phonon Raman spectrum. Experimentally, these results are most relevant in the low-temperature regime of the correlated insulating phase of TBG, where moir\'e phonons are expected to dominate the response and electronic contributions an be neglected \cite{Cao2018Correlated}.        
To examine how the phonon Raman spectrum evolves with the twist angle $\theta$, we compute the Raman intensity $\mathcal{I}(\omega)$ for various values of $\theta$ at fixed light polarization. The sum over phonon modes $n$ in Eq.~\eqref{eq:Raman_intensity} is reorganized as a sum over Raman-active irreducible representations $\Gamma$, and the corresponding modes $m$ within each $\Gamma$.
The intensity  now reads as
\begin{align}
	\mathcal{I}(\omega) \propto \sum_{\Gamma,m} \abs{ \sum_{\mu\nu} E_{\text{in}}^\mu \,  R^{\mu\nu}_{\Gamma,m} \, E_{\text{out}}^\nu } ^2 \, \frac{\gamma}{(\omega-\omega_{\Gamma,m})^2 + \gamma^2}. 
\end{align}
where $\gamma$ is the frequency-independent broadening arising from anharmonic effects, which we discuss in detail in SM \cite{suppl}. 
We work in the backscattering configuration with identical incident and scattered polarizations, $\bar{z}(xx)z$, where both $A_1$ and $E_2$ modes contribute to the Raman intensity $\mathcal{I}(\omega)$. The corresponding Raman tensors are provided in \cite{suppl}.
For comparison, we note that in the cross-polarized geometry $\bar{z}(xy)z$, only $E_2$ modes contribute. However, the qualitative features of the Raman response discussed below remain unchanged.

Figure~\ref{fig:ramanintensity}(a)–(c) shows the phonon Raman spectra for three twist angles $\theta$, focusing on frequencies below 80 meV, which is well below the lowest Raman-active phonon in the monolayer graphene \cite{ferrari2013}. For all angles, we find a series of peaks with intensities that generally decrease with frequency, consistent with the dominance of lower harmonics in the Raman tensor [Eq.~\eqref{eq:Raman_tensor}]. As $\theta$ decreases, additional peaks appear due to increased zone folding from the larger moiré supercell. Notably, modes within the same irreducible representation can have very different intensities, as seen in the weak peak near 10 meV in Fig.~2(c), which is symmetry-allowed but has low Raman weight.


Figure~\ref{fig:ramanintensity}(d) shows a color plot of the Raman spectra across a range of twist angles $\theta \in (0.85^\circ, 4^\circ)$ for frequencies up to 40 meV, where color indicates the intensity of the Raman response. A roughly linear dependence of the Raman-active frequencies on $\theta$ is observed, reflecting the folding of phonon modes in the moir\'e Brillouin zone. This dependence on the moir\'e length scale becomes more apparent when the frequency axis is rescaled by $\omega_0 = (2\pi/L_M)\sqrt{\lambda/\rho}$ \cite{koshino2019moire,ochoa2019moire}, with $L_M= a/(2 \sin(\theta/2))$, resulting in a data collapse shown in Fig.~\ref{fig:ramanintensity}(e). At small angles, where lattice reconstruction becomes more pronounced, the spectral gap between the lowest $A_1$ and $E_2$ modes exceeds the phonon linewidth $\gamma$, allowing these symmetry channels to be clearly resolved. This splitting is visible in both panels (d) and (e) for $\theta \lesssim 1^\circ$.

%
%

{\color{blue}\textit{Summary}}--
In this letter, we developed a microscopic theory
of Raman scattering from moir\'e phonons in TBG, focusing on the low-temperature correlated insulating regime where lattice degrees of freedom dominate the response. Moir\'e phonons, arising from interlayer sliding and stacking reconstruction include both acoustic-like phasons and a set of low-energy optical modes that are folded into the moir\'e Brillouin zone. We derived the Raman tensor by expanding the stacking-dependent polarizability and classified the Raman-active modes under the emergent $D_6$ symmetry of the moir\'e lattice. Our results reveal a sequence of low-frequency Raman peaks whose number increases and whose spacing decreases with twist angle, reflecting enhanced zone folding. Crucially, even modes belonging to the same irreducible representation can exhibit strongly varying Raman intensities, depending on the detailed structure of their phonon wavefunctions. This intensity variation provides additional information beyond symmetry alone and becomes especially pronounced at small angles, where lattice relaxation leads to a resolvable splitting between $A_1$ and $E_2$ modes. Our work establishes Raman spectroscopy as a sensitive probe of moir\'e phonons and lattice reconstruction in twisted van der Waals systems.

\begin{acknowledgments}
We thank Bernd Buechner, Rudolf Hackl, Eslam Khalaf, Rhine Samajdar and Jo{\~a}o A. Sobral for fruitful discussions. 
N.B.P.  and V.D. were supported by the U.S. Department   of Energy, Office of Science, Basic Energy Sciences under Award No. DE-SC0018056. N.B.P.  and V.D. acknowledge the hospitality of the Kavli Institute for Theoretical Physics (KITP) supported by grant NSF PHY-2309135. N.B.P. also acknowledges
the support  of the Alexander von Humboldt Foundation.
\end{acknowledgments}

\newpage
\renewcommand{\theequation}{S\arabic{equation}}\renewcommand{\thefigure}{S\arabic{figure}}
\setcounter{equation}{0}
\setcounter{figure}{0}

\begin{widetext}

\begin{center}
\bf{Supplemental information for ``Raman scattering from moiré phonons''}
\end{center}

\section{Moiré geometry: definitions and conventions}\label{sec: moire geom}
In this section we define the notation and conventions concerning the moiré geometry. Here we consider only the case of twisted bilayer graphene (TBG), but the whole formalism can be easily extended to other twisted van der Waals bilayers. 

First, let us define the convention for our lattice vectors on each graphene layer as $\textbf{a}_1 = a(1,0)$ and $\textbf{a}_2 = \frac{a}{2}(1,\sqrt{3})$, where $a \approx 2.46 \text{\AA}$ is the lattice constant. Therefore, the corresponding reciprocal lattice vectors are given by $\textbf{b}_i$, defined from the condition $\textbf{b}_i\cdot\textbf{a}_j = 2\pi \delta_{ij}$, which gives us $\textbf{b}_1 = \frac{2\pi}{a}(1,-1/\sqrt{3})$ and $\textbf{b}_2 = \frac{4\pi}{\sqrt{3}a}(0,1)$.  We construct the TBG lattice by taking two overlapping honeycomb lattices with AA stacking. Then, we rotate each layer by $\mp \theta/2$, as depicted in Fig. \ref{fig:figs1}(a). The rotation between the two layers can be performed by the $SO(2)$ transformation $\mathcal{R}_\theta$. The real and reciprocal lattice vectors in each layer ($l = 1,2$) are written in terms of the original vectors as: 
\begin{align}
	&\textbf{a}_i^{(l)} = \mathcal{R}_{\mp \theta/2}\,\textbf{a}_i  \qquad \textbf{b}_i^{(l)} = \mathcal{R}_{\mp \theta/2}\,\textbf{b}_i    \, ,
\end{align}
with $\mp$ for $l = 1,2$, respectively.

Locally, this operation can be identified as an  interlayer shift $\boldsymbol{\Delta}_0(\textbf{r})$ of an atom on layer 2 initially located at $\textbf{r}_0$ in the AA stacking to a new position $\textbf{r} = \mathcal{R}_\theta\, \textbf{r}_0$, as measured from layer 1\cite{koshino2019moire}:
\begin{align}\label{Delta_0}
	\boldsymbol{\Delta}_0(\textbf{r}) = \textbf{r} - \textbf{r}_0 = (1-\mathcal{R}_\theta^{-1})\textbf{r}
\end{align}
This allows us to define the primitive vectors of the moir\'e superlattice $\textbf{L}^M_i$ with the condition $\boldsymbol{\Delta}_0(\textbf{L}^M_i) = \textbf{a}_i$, that is, the maxima of the interference pattern coincide with the primitive vectors $a_i$ of the original AA-stacked bilayer. 
 Using the definition of $\boldsymbol{\Delta}_0(\textbf{r})$, we find that the
moir\'e lattice vectors are given by
 \begin{align}\textbf{L}_i^M = (1-\mathcal{R}^{-1}_\theta)^{-1}\textbf{a}_i^{(1)},
 \end{align} 
 which defines the moir\'e  unit cell. The period of the moir\'e  interference pattern is then
  $L_M  =\frac{a}{2\sin\pa{\theta/2}}$. The reciprocal space vectors of the moir\'e superlattice are denoted by $\textbf{G} = n_1\textbf{G}^M_1 + n_2\textbf{G}^M_2$, where the primitive vectors satisfy the condition $\textbf{G}^M_{i}\cdot \textbf{L}^{M}_j = 2\pi \delta_{ij}$, and are given by
\begin{align}
	\textbf{G}^{M}_i = (1-\mathcal{R}_\theta)\textbf{b}_i^{(1)} = \textbf{b}_i^{(1)} - \textbf{b}_i^{(2)}.
\end{align}
The vectors $\textbf{G}^M_i$ define the moiré Brillouin zone (mBZ), as shown in Fig. \ref{fig:figs1}(b). The high-symmetry points of the mBZ are denoted with a bar: $\overline{\Gamma}, \overline{M} , \overline{K} $ and $\overline{K}^\prime$. 

%


It is important to emphasize that the moir\'e beating pattern forms regardless of the choice of twisting center, but the point-group symmetry of the resulting superlattice depends on this choice.
 In this work, we consider the case where the rotation axis passes through a hexagon center, preserving the $\mathcal{C}_6$ symmetry of the monolayers (Fig.~\ref{fig:figs1}(c)).
 Additionally, there is a set of two-fold rotation axes on the same plane as the system. The layers are exchanged upon these transformations, and in a strictly 2D system they correspond to the  $\mathcal{C}_{2x}$ and $\mathcal{C}_{2y}$ rotations. 
 Together, these operations generate the point group $D_6$, which governs the symmetry of the moir\'e superlattice considered here.

\begin{figure}
	\centering
	\includegraphics[width=1\linewidth]{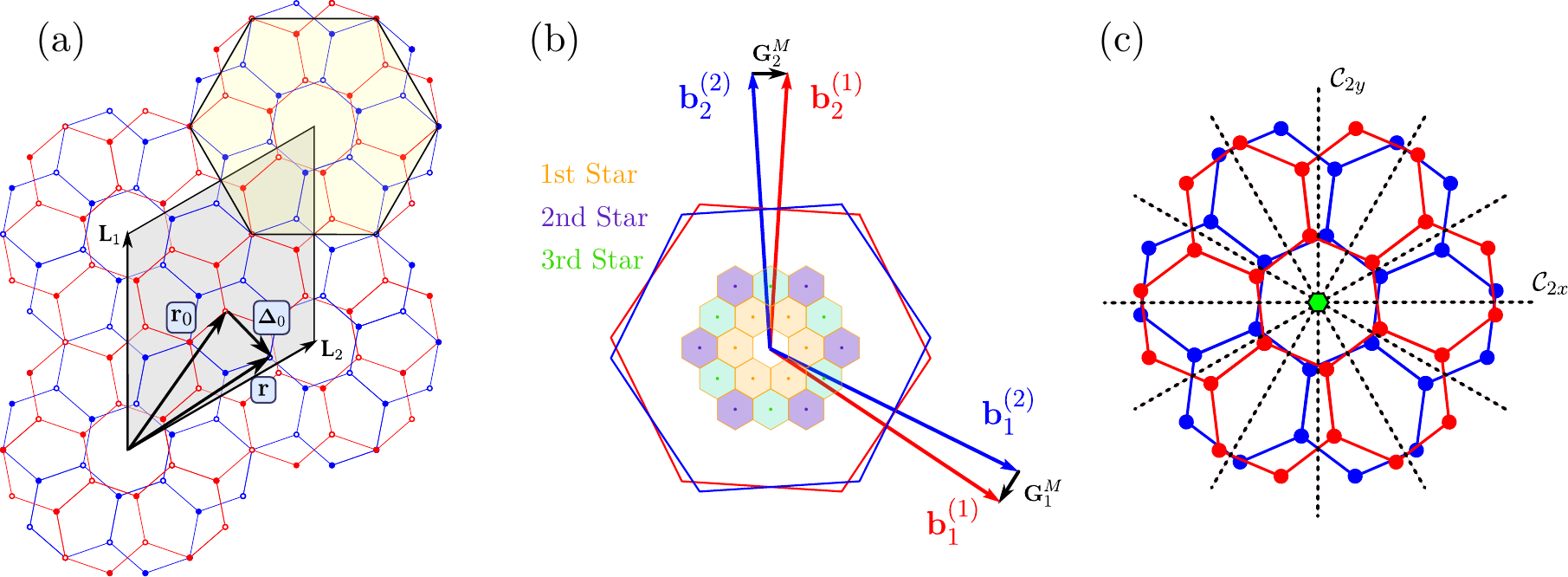}
	\caption{(a) TBG lattice structure. The moiré unit cell (mUC) is indicated by the gray area, and it is defined by the vectors $\textbf{L}_i^M$. The yellow area indicates the Wigner-Seitz unit cell in real space. The interlayer sliding vector $\boldsymbol{\Delta}_0$ is shown inside the mUC for one site. (b) Moiré Brillouin zone (mBZ) and reciprocal lattice vectors. The shaded areas correspond to the mBZ translated by $\textbf{G}$, where the different colors indicate the 3 first stars. (c) Symmetries of the TBG lattice. Here we choose the moiré rotation axis to coincide with a hexagon center,which preserves the 6-fold symmetry of the monolayer. The point-group is $D_6$, generated by the twofold rotations $\mathcal{C}_{2x}$ and $\mathcal{C}_{2y}$,  shown in by the dashed lines and the $\mathcal{C}_6$ rotation indicated by the green hexagon.}
	\label{fig:figs1}
\end{figure}

%
%
\section{Lattice relaxation }
On top of the rigid twist described by $\Delta_0(\textbf{r})$, the lattice sites are further displaced due to the interlayer adhesion potential $\mathcal{V}_{\text{ad}}$. This happens due to a competition between the adhesion energy, which favors locally stable AB stacking configurations, and the intralayer elastic energy, which resists deviations from the ideal lattice positions. As a result, the atomic structure relax towards a metastable configuration that minimizes the total elastic free energy $\mathcal{F}$. To obtain this relaxed structure, we closely follow the description outlined in Refs.\cite{koshino2019moire} and \cite{ochoa2019moire}. 

By including in-plane lattice deformations $\textbf{u}^{(l)}(\textbf{r})$, the sliding vector becomes:
\begin{align}
	\boldsymbol{\Delta}(\textbf{r}) = \Delta_0(\textbf{r}) + \textbf{u}^{(2)}(\textbf{r})-\textbf{u}^{(1)}(\textbf{r}).
\end{align}
The adhesion potential can be modeled as a periodic function of $\boldsymbol{\Delta}(\textbf{r})$, where its simplest form corresponds to the following 1st order Fourier expansion on the graphene reciprocal lattice:
\begin{align}\label{adhesion_pot}
	\mathcal{V}_{\text{ad}}[\textbf{r},\textbf{u}(\textbf{r})] = V_0 \sum_{j = 1}^3 \cos(\boldsymbol{\Delta}(\textbf{r}) \cdot \textbf{b}_j^{(1)} ) = V_0\sum_{j=1}^3 \cos \pa{ \textbf{G}^M_j \cdot \textbf{r} + \textbf{u}(\textbf{r})\cdot\textbf{b}_j^{(1)} }
\end{align}
where $\textbf{G}^M_3 = -\textbf{G}^M_1 - \textbf{G}^M_2$, $\textbf{b}^{(1)}_3 = -\textbf{b}^{(1)}_1 - \textbf{b}^{(1)}_2$, and we used the definition of $\boldsymbol{\Delta}_0$ in \eqref{Delta_0}. Also, we have introduced the \textit{relative} displacement field $\textbf{u}(\textbf{r}) \equiv \textbf{u}^{(2)}(\textbf{r})-\textbf{u}^{(1)}(\textbf{r})$. This form for $\mathcal{V}_{\text{ad}}$ is sometimes called the 1st star approximation, for reasons that will be clear later in this section. The parameter $V_0$, which gives the energy difference between AA and AB stacking configurations, is generally obtained through first-principle calculations, with different values ranging from $V_0 \sim 39 - 320$ meV/nm$^2$ \cite{zhou2015van,dai2016twisted,carr2018relaxation,samajdar2022moire}. Without lost of generality, we choose to set $V_0 \approx 90\, \text{meV/nm}^2$, as reported in Ref. \cite{carr2018relaxation}. 

The elastic free energy is composed by the contribution from each graphene layer, and it can be decomposed into a center-of-mass term that depends on the total displacement $\textbf{v}(\textbf{r}) =  \textbf{u}^{(2)}(\textbf{r}) + \textbf{u}^{(1)}(\textbf{r})$, and another that depends on the relative displacements $\textbf{u}(\textbf{r})$. Because the adhesion potential is a function of $\textbf{u}(\textbf{r})$ only, we can disregard the contributions from $\textbf{v}(\textbf{r})$, and the total energy becomes:
\begin{align}
	\mathcal{F}[\textbf{u}(\textbf{r})]  =  \int d^2\textbf{r}   \co{ \frac{\lambda}{2}(\nabla \textbf{u})^2 + \frac{\mu}{4}\pa{ \partial_\alpha u_\nu + \partial_\nu u_\alpha }^2   } + \int d^2\textbf{r} \,  \mathcal{V}_{\text{ad}}[\textbf{r},\textbf{u}^{}(\textbf{r})].
\end{align}
where the first term is the elastic energy corresponding to in-plane relative displacements $\textbf{u}(\textbf{r})$, and  $\lambda \approx 3.25 \text{eV/\AA}^2$ and $\mu \approx 9.57 \text{eV/\AA}^2$ are the Lamé coefficients of graphene \cite{zakharchenko2009finite,lee2008measurement}.  The optimal configuration is then obtained by minimizing the free energy above, $\delta \mathcal{F} = 0$, resulting in the following equations of motion:
\begin{align}
	\frac{\lambda+\mu}{2} \nabla \pa{ \nabla\cdot \textbf{u} }+ \frac{\mu}{2}\nabla^2\, \textbf{u} = \pdv{\textbf{u}} \mathcal{V}_{\text{ad}}[\textbf{r},\textbf{u}].
\end{align}
We assume the solution of the equation above, $\textbf{u}_0(\textbf{r})$, to be periodic with the moiré pattern: $\textbf{u}_0(\textbf{r}) = \sum_\textbf{G} e^{i \textbf{G} \cdot \textbf{r}} \,  \textbf{u}_\textbf{G}$, and the differential equation is reduced to a set of self-consistent algebraic equations in the moiré reciprocal lattice:
\begin{align}\label{self-consist_1}
	&\mathbb{K}_\textbf{G} \textbf{u}_\textbf{G} = 2V_0\sum_{j=1}^3 f^j_\textbf{G} \textbf{b}_j^{(1)}, \\
	&\label{self-consist_2} f_\textbf{G}^j = \frac{1}{A_m}\int d^2\textbf{r}\,\,e^{-i\textbf{G}\cdot\textbf{r}} \sin\pa{ \textbf{G}^M_j \cdot \textbf{r} + \urz\cdot\textbf{b}_j^{(1)} },
\end{align}		
where $f^j_\textbf{G}$ is the inverse Fourier transform of the force terms in the equation of motion, and  $\mathbb{K}_\textbf{G}$ is the kinetic term written as a $2\cross2$ matrix:
\begin{align}\label{K_G}
	\mathbb{K}_\textbf{G} = 
	\begin{pmatrix}
		(\lambda+2\mu)G_x^2 + \mu G_y^2 &&	(\lambda + \mu )G_xG_y\\
		(\lambda + \mu )G_xG_y && 	(\lambda+2\mu)G_y^2 + \mu G_x^2
	\end{pmatrix},
\end{align}

\subsection{Symmetry constraints}

The self-consistent problem outlined above can be addressed using various numerical methods, such as gradient descent, which has proven effective in moiré systems at very small twist angles \cite{ramos2025flat}. In our case, however, it is enough to employ a simple fixed-point iteration scheme. In order to obtain a solution that is not only periodic in the moiré scale, but also respects all the symmetries of the underlying point-group, we define the \textit{stars} of moiré reciprocal vectors $\textbf{G}$. Each star consists of all symmetry-equivalent $\textbf{G}$ vectors obtained by applying the rotations and reflections of $D_6$, where a representative value is denoted by  $\textbf{G}^{(\alpha)}$. We show a set of stars in Fig.\ref{fig:figs1}(c). 

The transformation laws of the vector field $\textbf{u}_0(\textbf{r})$ requires that $\textbf{u}_\textbf{G}$ transforms under in and out-of-plane generators of $D_6$ as
\begin{align}
	\mathcal{C}_{6}\textbf{u}_{\textbf{G}} = \textbf{u}_{\mathcal{C}_6\textbf{G}} \qquad 	\mathcal{C}_{2y}\textbf{u}_{\textbf{G}} = -\textbf{u}_{\mathcal{C}_{2y}\textbf{G}} 
\end{align}
Therefore, by expanding the displacement field in terms of the stars $\textbf{G}^{(\alpha)}$ together with the relations above, we can solve the set of equations \eqref{self-consist_1}-\eqref{self-consist_2} with a reduced number of degrees of freedom while preserving the point-group symmetries.  We also point out that previous works have shown that a small number of stars close to the $\Gamma$ point is enough to obtain the relaxed structure \cite{koshino2019moire,nam2017lattice,ochoa2019moire,ochoa2022degradation}. This comes from the fact that $|\textbf{u}_\textbf{G}|$ decays rapidly for higher harmonics. In all of our numerical calculations we use up to 6 momentum shells, which corresponds to $N_{\text{star}} = 21$, and less stars ($N_\text{star} \leq 6$) are used for large angles ($\theta \gtrsim 3^\circ$). 


%
%
\section{Moiré phonons }

Now we consider fluctuations on top of the obtained relaxed structure: $\delta \textbf{u}(\textbf{r},t) = \textbf{u}(\textbf{r},t) - \textbf{u}_0(\textbf{r},t) $. The free energy in the harmonic approximation is then given by two non-zero contributions, $\mathcal{F}_0$, corresponding to the equilibrium solution, and the quadratic term in the fluctuations, $\mathcal{F}^{(2)}[\delta\textbf{u}(\textbf{r})]$:
\begin{align}
	\mathcal{F}[\delta\textbf{u}(\textbf{r})] = \mathcal{F}_0 +  \int d^2\textbf{r} \, \co{ \frac{\lambda}{4}\pa{\nabla \cdot \delta\textbf{u}}^2 + 
		\frac{\mu}{4}\pa{\partial_i\delta u_j + \partial_j\delta u_i}^2 + \left. 
		\frac{1}{2!} \frac{\partial^2 \mathcal{V}_\text{ad}}{\partial u_i\partial u_j}\right|_0 \, \delta u_i \delta u_j }.
\end{align}
The total Hamiltonian  of the system is then $\mathcal{H} = \mathcal{T} + \mathcal{F}$, where $\mathcal{T} = \int d^2r \rho^{-1} \delta \boldsymbol{\pi}^2(\textbf{r})$, where $\rho = 7.6 \times 10^{-7}$ kg/m$^2$ is the mass density of graphene and $\delta \boldsymbol{\pi}(\textbf{r})$ is the canonical momentum conjugated to the field $\delta \textbf{u}(\textbf{r})$. The moiré phonon field admits a Fourier expansion like
\begin{align}
	\delta \textbf{u}(\textbf{r},t) = \sum_\textbf{q} e^{i\textbf{q}\cdot \textbf{r}} \delta \textbf{u}_\textbf{q}(t) = \sum_{\textbf{q} \in \text{mBZ}}\sum_\textbf{G} e^{i(\textbf{q}+\textbf{G})\cdot\textbf{r}}\delta \textbf{u}_{\textbf{q}+\textbf{G}}(t) \nonumber, 
\end{align}
with a similar expansion for $\delta \boldsymbol{\pi}(\textbf{r})$, so the total Hamiltonian in momentum space is written as:
\begin{align}\label{Hamiltonian_harmonic}
	\mathcal{H} = \sum_{\textbf{q} \in \text{mBZ}}\sum_\textbf{G} \frac{1}{\rho} \delta\boldsymbol{\pi}_{\textbf{q}+\textbf{G}}^\dagger \cdot \delta\boldsymbol{\pi}_{\textbf{q}+\textbf{G}} + \frac{1}{2}\sum_{\textbf{q} \in \text{mBZ}}\sum_{\textbf{G},\textbf{G}'} \delta \textbf{u}_{\textbf{q}+\textbf{G}'}^\dagger \mathbb{D}_{\textbf{q}}(\textbf{G},\textbf{G}') \, \delta \textbf{u}_{\textbf{q}+\textbf{G}}
\end{align}
where we introduced the dynamical matrix $\mathbb{D}_\textbf{q}$. For every $\textbf{q}\in \text{mBZ}$ this is a $2N_G$--dimensional matrix composed by $2\times 2$ blocks in the basis of $\textbf{G}$ vectors:
\begin{align}
	\mathbb{D}_\textbf{q}(\textbf{G},\textbf{G}') =  \mathbb{K}_\textbf{q+\textbf{G}}\delta_{\textbf{G}\textbf{G}'} + 2\mathbb{V}_{\textbf{G}-\textbf{G}'}.
\end{align}
The matrix  $\mathbb{K}_\textbf{q}$ is given by Eq. \eqref{K_G}, and $\mathbb{V}_\textbf{G}$ is $2\times 2$ matrix defined as
\begin{align}
	\mathbb{V}_\textbf{G} = \frac{1}{A_m}\int d^2\textbf{r}\,\, e^{-i\textbf{G}\cdot \textbf{r}} \left. 
	\frac{\partial^2 \mathcal{V}_\text{ad}}{\partial u_i\partial u_j}\right|_0 
	= -\frac{V_0}{A_m}\sum_{j=1}^{3}
	\int d^2\textbf{r} \,\,e^{-i\textbf{G}\cdot \textbf{r}} \cos \pa{ \textbf{G}^M_j \cdot \textbf{r} + \textbf{u}_0(\textbf{r})\cdot\textbf{b}_j^{(1)} }	\begin{pmatrix}
		b_{j,x}^2 & b_{j,x}b_{j,y}\\
		b_{j,y}b_{j,x} & b_{j,y}^2\\
	\end{pmatrix} 
\end{align}
where we used the explicit form of the adhesion potential in Eq. \eqref{adhesion_pot}. Our job now is to diagonalize the dynamical matrix, which satisfies the secular equation 
\begin{align}
	\sum_{\textbf{G}'} \mathbb{D}_{\textbf{q}}(\textbf{G},\textbf{G}')  \textbf{C}_{n,\textbf{q}}(\textbf{G}') = \rho\omega_{n,\textbf{q}}^2\textbf{C}_{n,\textbf{q}}(\textbf{G}).
\end{align}	
This can be derived from the equations of motion obtained from the Hamiltonian in Eq. \eqref{Hamiltonian_harmonic}. Here the index $n$ refers to the moiré phonon branch, and $\textbf{C}_{n,\textbf{q}}(\textbf{G})$ is a set of eigenvectors satisfying the normalization condition $\sum_\textbf{G}\abs{C_{n,\textbf{q}}(\textbf{G})}^2 = 1$. Then, for a given mode $\omega_{n,\textbf{q}}$ the eigenvectors can be organized as a $2N_G$ component array
\begin{align}
	\boldsymbol{\xi}_{n,\textbf{q}}^T = \co{ \textbf{C}_{n,\textbf{q}}(\textbf{G}_0) \, , \,  \textbf{C}_{n,\textbf{q}}(\textbf{G}_1) \, , \,  \cdots, \textbf{C}_{n,\textbf{q}}(\textbf{G}_{N_G}) },
\end{align}
such that $	\mathbb{D}_\textbf{q} \, \boldsymbol{\xi}_{n,\textbf{q}} = \rho\omega_n^2 \boldsymbol{\xi}_{n,\textbf{q}}$. Therefore, we just have to numerically diagonalize $\mathbb{D}_\textbf{q}$ for all moiré reciprocal lattice points $\{\textbf{G}_i\}$ within a given set of stars $\textbf{G}^{(\alpha)}$. The moiré phonon dispersion is obtained along the high-symmetry path $\bar{K}\bar{\Gamma}\bar{M}\bar{K}'$, as shown in Fig. 1 of the main text. For $\textbf{q}=0$, the modes $\boldsymbol{\xi}_{n,\textbf{q}=0}$ can be classified into irreps of $D_6$, as outlined in in the main text.

Once $\mathbb{D}_\textbf{q}$ is diagonalized, the normalized eigenvectors $\textbf{C}_{n,\textbf{q}}(\textbf{G})$ provide a basis for a normal mode expansion:
\begin{align}\label{normal_modes_expan}
	&\delta \uv{\textbf{G}+\textbf{q}} = \sum_n \textbf{C}_{n,\textbf{q}}(\textbf{G}) Q_{n,\textbf{q}} \qquad \delta \boldsymbol{\pi}_{\textbf{G}+\textbf{q}} = \sum_n \textbf{C}^{*}_{n,\textbf{q}}(\textbf{G}) P_{n,\textbf{q}}.
\end{align}
The next step is to promote the normal coordinates into operators $\co{ Q_{n,\textbf{q}} ,	\, P_{n^\prime,\textbf{q}^\prime} } = i\hbar \delta_{nn'}\delta_{\textbf{q},\textbf{q}'}$, where we introduce the creation and annihilation operators through:
\begin{align}\label{QandP_normalmodes}
	&Q_{n,\textbf{q}} = \sqrt{\frac{\hbar}{2\rho \omega_{n,\textbf{q}}}}\, \pa{ a_{n,-\textbf{q}}^\dagger + a_{n,\textbf{q}} } 
	\qquad 
	P_{n,\textbf{q}} = \sqrt{\frac{\hbar\rho \omega_{n,\textbf{q}}}{2}}\, \pa{ a_{n,-\textbf{q}}^\dagger - a_{n,\textbf{q}} },
\end{align}
and then, the phonon hamiltonian takes the diagonal form:
\begin{align}
	\mathcal{H}_{\text{ph}} &= \sum_{\textbf{q}\in \text{mBZ}} \sum_n \frac{1}{2\rho} P^\dagger_{n,\textbf{q}} P_{n,\textbf{q}} + \frac{\rho}{2}\omega_{n,\textbf{q}} Q_{n,\textbf{q}}^\dagger Q_{n,\textbf{q}} \\ 
	& = \sum_{\textbf{q}\in \text{mBZ}} \sum_n \hbar\omega_{n,\textbf{q}} \pa{a_{n,\textbf{q}} a^\dagger_{n,\textbf{q}} + 1/2  }
\end{align}
where we used $[a_{n,\textbf{q}}, a^\dagger_{n',\textbf{q}'}] = \delta_{nn'}\delta_{\textbf{q},\textbf{q}'}$.  More importantly, the displacement field $\delta \textbf{u}(\textbf{r})$ is written in terms of $a_{n,\textbf{q}}$ and $a^\dagger_{n,\textbf{q}}$ as:
\begin{align} \label{u_Fouriertransf}
	\delta \textbf{u}(\textbf{r}) &= \frac{1}{\sqrt{A}} \sum_{\textbf{q}\in \text{mBZ}}\sum_\textbf{G}  \delta\textbf{u}_{\textbf{G}+\textbf{q}}\,  e^{i\pa{\textbf{G} + \textbf{q}}\cdot \textbf{r}} \\
	& = \frac{1}{\sqrt{A}} \sum_{\textbf{q}\in \text{mBZ}}\sum_{\textbf{G}, n} \textbf{C}_{n,\textbf{q}}(\textbf{G}) \,  e^{i\pa{\textbf{G} + \textbf{q}}\cdot \textbf{r}} \times \sqrt{\frac{\hbar}{2\rho\omega_{n,\textbf{q}}} }  \pa{a_{n,\textbf{q}} + a_{n,-\textbf{q}}^\dagger }
\end{align}

Because we will be interested in the $\textbf{q} = 0$ modes in the Raman scattering calculation, it is enough to write the displacement field as:
\begin{align}
	\delta \textbf{u}(\textbf{r}) =  \frac{1}{\sqrt{A}} \sum_{\textbf{G},n} \textbf{C}_{n}(\textbf{G}) \,  e^{i\textbf{G}\cdot \textbf{r}} \times \sqrt{\frac{\hbar}{2\rho\omega_{n}} }  \pa{a_{n} + a_{n}^\dagger }
\end{align}

%
%

\section{Raman intensity}

We wish to compute the Raman intensity $I(\Omega)$ contribution from the moiré phonons. This quantity is defined as the time-ordered correlation function of the Raman operator $\mathcal{R}(t) = \sum_{\mu\nu}E_{\text{in}}^\mu E_{\text{out}}^\nu \,  \chi^{\mu\nu}_{\text{Ram.}}(t):$
\begin{align}
	I(\Omega) = \int dt e^{i\Omega t} \expval{T\, \mathcal{R}(t)\, \mathcal{R}(0)}.
\end{align}
The Raman scattering only takes place if the polarizability of the medium $\chi^{\mu\nu}$ changes due to the excitations in the crystal. In our case, this corresponds to the moiré phonon contribution, so we may expand the polarizability in terms of the moiré phonon field $\delta u_\lambda(\textbf{r},t)$, where the 1st order term corresponding to the Raman scattering in the continuum approximation reads as
\begin{align}
	\chi^{\mu\nu}_{\text{Ram.}}(t) 
	& = \int d^2\mathbf{r}\,\, \widetilde{R}^{\mu\nu}_{\lambda}(\textbf{r}) \,\delta u_\lambda(\textbf{r}, t),
\end{align}
where the integral is over a moir\'e cell of area $A_m$ and $\lambda$ labels in-plane Cartesian coordinates $x,y$ (hereafter repeated Greek indices are summed). The rank-3 Raman tensor field $\widetilde{R}^{\mu\nu}_{\lambda}(\textbf{r})$ accounts for the modulations of $\chi^{\mu\nu}$ in the stacking order fluctuations $\delta \textbf{u}(\textbf{r},t)$, and it is assumed to be smooth on the microscopic scale $a$, the original lattice constant of each individual graphene layer.  More specifically, $\widetilde{R}^{\mu\nu}_{\lambda}(\textbf{r})$ is defined as the derivative of the polarizability tensor with respect to the fluctuation evaluated at equilibrium,  
\begin{align}
	\label{R(r)definition}
	\left. \widetilde{R}^{\mu\nu}_{\lambda}(\textbf{r})\equiv \frac{\partial \chi^{\mu\nu}[\textbf{u}(\textbf{r})]}{\partial u_{\lambda}}\right|_{\textbf{u}=\textbf{u}_0(\mathbf{r})},
\end{align}
where now $\chi^{\mu\nu}[\textbf{u}(\textbf{r})]$ is a stacking-dependent polarizability tensor of the bilayer.

Before discussing the detailed form of the polarizability tensor, it is useful to show that the Raman operator $\mathcal{R}(t)$ can be expressed in terms of the normal coordinates $Q_n(t)$ and a position-independent Raman tensor $R^{\mu\nu}_n$, whose components are determined by the symmetry of the moiré superlattice in analogy to the conventional theory of light scattering in crystals 
\cite{hayes1978scattering,poulet1970spectres}.
We thus make use of the expansion in Eq. \eqref{normal_modes_expan} to write:
\begin{align}
	\chi^{\mu\nu}_{\text{Ram.}}(t) =  
	\sum_{\textbf{G},n} \widetilde{R}^{\mu\nu}_{\lambda}(-\textbf{G}) C^\lambda_{n}(\textbf{G}) \,  Q_n(t), 
\end{align}
where we defined the Fourier components of the Raman tensor field as
\begin{align}\label{R(r)Fouriertrans}
	\widetilde{R}^{\mu\nu}_\lambda(\textbf{G}) \equiv \frac{1}{A_m}\int d^2\mathbf{r} \,\, \widetilde{R}^{\mu\nu}_{\lambda}(\textbf{r}) e^{-i\textbf{G}\cdot \textbf{r}}.
\end{align}

Next, we perform a variable transformation to write the local rank-3 tensor $\widetilde{R}^{\mu\nu}_\lambda(\textbf{r}_i)$ in the basis of normal coordinates:
\begin{align}\label{Rr_toQ}
	\widetilde{R}_\lambda^{\mu\nu}(\textbf{r}) = \sum_n\co{ \pdv{Q_n}{u_\lambda(\textbf{r})}\pdv{\chi^{\mu\nu} }{Q_n} }_0 = \sum_n\co{\pdv{Q_n}{u_\lambda(\textbf{r})}}_0  \,R^{\mu\nu}_n,
\end{align}
where we defined the proper \textit{Raman tensor} from the moiré superlattice as $R_n^{\mu\nu} \equiv (\partial \chi^{\mu\nu}/\partial Q_n)_0$, which must not be confused with its counterpart obtained from the graphene lattice vibrations. The relative displacements $\delta u_\lambda(\textbf{r})$ are linear in $Q_n$, so we have
\begin{align}
	Q_n = \sum_\textbf{G} \frac{1}{A_m}\int d^2\textbf{r} \,  e^{-i\textbf{G}\cdot\textbf{r}} \co{ C^\rho_n(\textbf{G}) }^* \delta u_\rho(\textbf{r}).
\end{align}
The Jacobian of this transformation is
\begin{align}
		\pdv{Q_n}{u_\lambda(\textbf{r})} &= 
		\sum_\textbf{G}\frac{1}{A_m}\int d^2\textbf{r}' \, e^{-i\textbf{G}\cdot\textbf{r}^\prime}\co{ C^\rho_n(\textbf{G}) }^* \delta_{\lambda\rho}\delta\pa{\textbf{r} - \textbf{r}'}+
		\sum_\textbf{G} \frac{1}{A_m}\int d^2\textbf{r}' \, \pdv{e^{-i\textbf{G}\cdot\textbf{r}'}}{u_\lambda(\textbf{r})}\co{ C^\rho_n(\textbf{G}) }^* \delta u_\rho(\textbf{r}').
	\end{align}
The second term vanishes. Since $Q_n$ and $\delta u_\lambda(\textbf{r})$ are treated as independent fields, the exponential has no implicit dependence on $u_\lambda(\textbf{r})$. Even if such dependence is assumed, the derivative is evaluated at equilibrium, where $\delta u_\lambda(\textbf{r}) \to 0$. Both arguments are consistent with the linear response regime used to define $\mathcal{R}(t)$.

Therefore, we have only the first term, and the position dependent tensor takes the form
\begin{align}
	\widetilde{R}^{\mu\nu}_\lambda(\textbf{r}) = 		\sum_{\textbf{G} , n} e^{-i\textbf{G}\cdot\textbf{r}}\co{ C^\lambda_n(\textbf{G}) }^*R^{\mu\nu}_n, \,
\end{align}
while the Fourier component $\widetilde{R}^{\mu\nu}_\lambda(\textbf{G})$ is simply
\begin{align}\label{R(G)-to-Rn}
	R^{\mu\nu}_{\lambda}(-\textbf{G}) = \sum_n \co{ C^\lambda_n(\textbf{G}) }^* R^{\mu\nu}_n.
\end{align}
Finally, this contribution to the Raman polarizability is:
\begin{align}
	\chi^{\mu\nu}_{\text{Ram}} = \sum_{\textbf{G},n,n'} \co{ C^\lambda_{n'}(\textbf{G}) }^* C^\lambda_n(\textbf{G}) R^{\mu\nu}_n Q_n(t) = \sum_n R^{\mu\nu}_n \, Q_n(t).
\end{align}
Therefore, there is no explicit interference effect from the $\textbf{G}$ terms, and the Raman intensity retains the same functional form as in a periodic crystal, despite the spatial dependence of $\chi^{\mu\nu}[\textbf{u}(\textbf{r})]$. 
The key difference lies in the fact that $Q_n$ corresponds to optical modes of the moiré phonons, not of the decoupled layers.

After projecting the modes onto irreps $\Gamma$ of the $D_6$ point group, the Raman intensity takes the form:
\begin{align}\label{Raman_intensity}
	\mathcal{I}(\Omega) = 2\co{ n_B(\Omega)+1}& \sum_{\Gamma,m} \abs{ \sum_{\mu\nu} E_{\text{in}}^\mu \,  R^{\mu\nu}_{\Gamma,m} \, E_{\text{out}}^\nu }^2 \Im D^R_{\Gamma,m}(\Omega, \textbf{q}=0), 
\end{align}
where $D^R_{\Gamma,m}(\Omega, \textbf{q}=0)$ is the retarded phonon Green's function of a $\textbf{q}=0$ mode $m$ transforming according to the irrep $\Gamma$. For in-plane vibrations, there are only two Raman-active irreps in the $D_6$ group, the symmetric $A_1$ mode, and the doublet $E_2$. The Raman tensors corresponding to these irreps are generically given --in the crystallographic basis-- by
\begin{align}
	R_{A_1} = 
	\begin{pmatrix}
		a & 0  \\
		0 & a  \\
	\end{pmatrix} 
	\quad 
	R_{E_2,1} = 	\begin{pmatrix}
		d & 0  \\
		0 & -d \\
	\end{pmatrix} 
	\, ,
	R_{E_2,2} = 	\begin{pmatrix}
		0 & -d \\
		-d & 0 \\
	\end{pmatrix} 	,
\end{align}
where $a$ and $d$ are phenomenological constants. Notice that distinct modes transforming according to the same irrep can have different values of constants, as we are going to see in the next section.

\subsection{Stacking-dependent polarizability}

Now we turn to the explicit dependence of the polarizability with the stacking configuration $\textbf{u}(\textbf{r})$. The basic idea is that the Raman tensor field can be understood as a functional on stacking configurations evaluated at the equilibrium configuration of the bilayer, $\widetilde{R}^{\mu\nu}_\lambda(\mathbf{r})=\widetilde{R}^{\mu\nu}_\lambda[\textbf{u}_0(\mathbf{r})]$, as defined in Eq. \eqref{R(r)definition}. We do not know this functional, but we can make a guess based on symmetry noting that, by construction, $\chi^{\mu\nu}[\textbf{u}]$ must be periodic with vectors $\textbf{b}_j$ of the reciprocal lattice of a commensurate bilayer. For in-plane components, the ansatz for modulations in the moiré reciprocal space is of the form
\begin{align}
	\chi^{\mu\nu}[\textbf{u}]\sim \sum_{s=1}^{N_\text{star}}\sum_{j=1}^3\left[\hat{\mathbf{G}}^s_j\right]^{\mu} \left[\hat{\mathbf{G}}^s_j\right]^{\nu}\,\cos  \pa{\textbf{G}^s_j \cdot \textbf{r} + \textbf{u}(\textbf{r})\cdot\textbf{b}_j^s }
	\,\,\,\,\textrm{for}\,\, \mu,\nu=x,y.
\end{align}
Here the upper index $s$ indicates the star for both the moiré and graphene reciprocal lattice vectors $\textbf{G}_j$ and $\textbf{b}_j$, as defined in Section \ref{sec: moire geom}. The unit vectors $\hat{\textbf{G}}^s_j=\textbf{G}^s_j/|\textbf{G}^s_j| $ are introduced to ensure that $\chi^{\mu\nu}$ respects the point group symmetry. Then, taking the derivative with respect to stacking fields, we end up with the following parametrization of the components of the stacking-dependent Raman tensor:
\begin{align}
\widetilde{R}^{\mu\nu}_{\lambda}[\textbf{u}]=\sum_{s=1}^{N_\text{star}}\zeta_s \, \sum_{j=1}^3 \,\,
\left[\hat{\textbf{b}}^s_j\right]_{\lambda}\left[\hat{\mathbf{G}}^s_j\right]^{\mu} \left[\hat{\mathbf{G}}^s_j\right]^{\nu}
 \,\sin \pa{\textbf{G}^s_j \cdot \textbf{r} + \textbf{u}(\textbf{r})\cdot\textbf{b}_j^s }
 \,\,\textrm{for}\,\, \mu,\nu=x,y,
\end{align}
where we introduced the dimensionless constants $\zeta_s$. These constants are defined as the relative intensity between the Raman response of a mode $n$ with some reference value. In the same manner as the adhesion potential, it is reasonable to assume that the lowest harmonics contribute more to the polarizability. Hence, we can generically assume $\zeta_s$ as a decaying function of the star $s$: $\zeta_s \sim \abs{\textbf{G}^M_j}/\abs{\textbf{G}^s_j}$, where $\textbf{G}^{M}_j$ is equivalent to the 1st star vectors. This ultimately leads to the decaying of Raman intensities for large frequencies, as shown in Fig 2 of the main text. 

Finally, by using Eq. \eqref{R(r)Fouriertrans} and the inverse of Eq. \eqref{R(G)-to-Rn}, we obtain the explicit form of the Raman tensor components in terms of the eigenmodes $\textbf{C}_n(\textbf{G})$:
\begin{align}
	R^{\mu\nu}_n = \sum_{s=1}^{N_\text{star}}\sum_{\textbf{G}}\sum_{j=1}^3 \zeta_s \,\, C^\lambda_n(\textbf{G})\,  \left[\hat{\textbf{b}}^{s}_j\right]_{\lambda}\left[\hat{\mathbf{G}}^s_j\right]^{\mu} \left[\hat{\mathbf{G}}^s_j\right]^{\nu} f_j^s(\textbf{G}),
\end{align}
where $f_j^s(\textbf{G})$ is the Fourier transform:
\begin{align}
	f^s_j(\textbf{G}) = \frac{1}{A_m}\int d^2\textbf{r}\,  e^{-i\textbf{G}\cdot\textbf{r}} \,\sin \pa{\textbf{G}^s_j \cdot \textbf{r} + \textbf{u}(\textbf{r})\cdot\textbf{b}_j^s }
\end{align}

%
%

%
%
\section{Anharmonicity effects}
To obtain a more realistic description of the Raman spectrum of moiré phonons we need to compute the phonon linewidth $\gamma$ due to anharmonic interactions. The linewidth is directly related to the imaginary part of the phonon self-energy $\Pi(\q,\Omega_{n,\q})$ and determines the inverse lifetime $\tau^{-1}$ of the moiré phonons. 

Here, we focus on the lowest-order contribution beyond the harmonic approximation, which corresponds to three-phonon scattering  processes.  This gives rise to the following contribution to the stacking free energy:
\begin{align}
	\mathcal{F}[\textbf{u}(\textbf{r})] = \mathcal{F}_0	+ \mathcal{F}^{(2)} + \int d^2 \textbf{r} \,\, \frac{1}{3!} \co{ \frac{\partial^3 \mathcal{V}_\text{ad}}{\partial u_\alpha \partial u_\beta \partial u_\gamma}}_0 \,\, \delta u_\alpha(\textbf{r}) \delta u_\beta(\textbf{r}) \delta u_\gamma(\textbf{r}),
\end{align} 
where $\mathcal{F}_0$ is the relaxed structure free energy and the second term corresponds to the quadratic terms in the stacking fluctuation $\delta\textbf{u}(\textbf{r})$, from which we obtain the dynamical matrix and the moiré phonon modes. Then, the last term can be treated as a perturbation $\mathcal{H}_{\text{int}}$ on top of the free Hamiltonian.

In momentum space this term is given by:
\begin{align}\label{Hint_q}
	\mathcal{H}_{\text{int}}[\textbf{u}(\textbf{r})] = \frac{1}{3!} \sum_{\textbf{q}_1,\textbf{G}_1}\sum_{\textbf{q}_2,\textbf{G}_2}\sum_{\textbf{q}_3,\textbf{G}_3} \mathcal{V}^{(3)}_{\alpha\beta\gamma}(\q_1+ \textbf{G}_1 + \q_2+ \textbf{G}_2 + \q_3+ \textbf{G}_3) \delta u^{\alpha}_{\textbf{G}_1+\q_1} \delta u^{\beta}_{\textbf{G}_2+\q_2} \delta u^{\gamma}_{\textbf{G}_3+\q_3}
\end{align}
where we used the Fourier expansion of $\delta \textbf{u}(\textbf{r})$ in Eq. \ref{u_Fouriertransf}, and we defined:
\begin{align}
	\mathcal{V}^{(3)}_{\alpha\beta\gamma}(\textbf{q}) = \frac{1}{A^{3/2}}\int d^2\textbf{r}\,\, e^{i\textbf{q}\cdot\textbf{r}} \co{ \frac{\partial^3 \mathcal{V}_\text{ad}}{\partial u_\alpha \partial u_\beta \partial u_\gamma}}_0.
\end{align}

We can further simplify the expression in Eq. \eqref{Hint_q} by writing it in terms of the normal modes defined by Eq. \eqref{u_Fouriertransf}:
\begin{align}
	\mathcal{H}_{\text{int}}[\textbf{u}(\textbf{r})] = &\frac{1}{3!} \sum_{\{n,\textbf{q},\textbf{G} \}} \mathcal{V}^{(3)}_{\alpha\beta\gamma}(\q_1+ \textbf{G}_1 + \q_2+ \textbf{G}_2 + \q_3+ \textbf{G}_3)\, \times \\
	&  \times C^{\alpha}_{n_1,\q_1}(\textbf{G}_1)C^{\beta}_{n_2,\q_2}(\textbf{G}_2)C^{\gamma}_{n_3,\q_3}(\textbf{G}_3)\,   \hat{Q}_{n_1,\q_1}\hat{Q}_{n_2,\q_2}\hat{Q}_{n_3,\q_3},
\end{align}
where the set $\{n,\textbf{q},\textbf{G} \}$ corresponds to all indices from 1 to 3. In principle this expression would be sufficient to write the three-phonon vertex interaction. However, it is convenient to introduce the phonon field operator $A_{n,\textbf{q}} = \pa{a_{n,\textbf{q}} + a^\dagger_{n,-\textbf{q}}} $, so we can absorb all the normalization factors in $Q_{n,\q}$ inside the vertex. The interaction Hamiltonian is then
\begin{align}
	\mathcal{H}_{\text{int}} = 
	\sum_{\{n,\textbf{q}\}} \Lambda^{n_1,n_2,n_3}_{\q_1,\q_2,\q_3} \, \hat{A}_{n_1,\q_1} \hat{A}_{n_2,\q_2} \hat{A}_{n_3,\q_3},
\end{align}
where the three-phonon vertex is defined as
\begin{align}
	\Lambda^{n_1,n_2,n_3}_{\q_1,\q_2,\q_3} = 
	\frac{1}{3!}\pa{\frac{\hbar}{2\rho}}^{3/2}\frac{1}{\sqrt{\prod_{j=1}^{3} \Omega_{n_j,\q_j} }} \, 
	\sum_{\{\textbf{G}\}} & \, \mathcal{V}^{(3)}_{\alpha\beta\gamma}\pa{\q_1+ \textbf{G}_1 + \q_2+ \textbf{G}_2 + \q_3+ \textbf{G}_3} \times \nonumber \\ 
	& \times C^{\alpha}_{n_1,\q_1}(\textbf{G}_1)C^{\beta}_{n_2,\q_2}(\textbf{G}_2)C^{\gamma}_{n_3,\q_3}(\textbf{G}_3).
\end{align}
Here we used the explicit form of $\hat{Q}_{n,\q}$ as in Eq. \eqref{QandP_normalmodes} where $\Omega_{n,\q}$ is the moiré phonon frequency. This interaction allows a phonon at a given momentum to decay into two lower-energy phonons or, conversely, for two phonons to combine into a single higher-energy mode, leading to a broadening of the Raman spectra.

\subsection{Moiré-phonon Self-energy}

The goal now is to compute the dressed phonon Green's function $D_n(\textbf{q},\Omega)$, which is related to the phonon self-energy $\Pi_n(\textbf{q},\Omega)$ through Dyson equation:
\begin{align}
	D^{-1}_n(\q,\Omega) =  \co{D_n^{(0)}(\q,\Omega) }^{-1} - \Pi_n(\textbf{q},\Omega),
\end{align}
where $n$ indicates the phonon branch and $D_n^{(0)}(\q,\Omega)$ is the bare retarded Green's function. To perform this calculation at finite temperature, it is convenient to switch to the imaginary-time formalism by introducing $\tau = it$ and expressing the propagators in terms of Matsubara frequencies. In this representation, the bare propagator is given by:
\begin{align}
	\mathcal{D}_{n}^{(0)}(\textbf{q},i\Omega_m) &= -\frac{1}{\beta} \int_{0}^{\beta} d\tau \, e^{i\Omega_m\tau}\,  \expval{T_\tau \, A_{n,\textbf{q}}(\tau) A^\dagger_{n,\textbf{q}}(0)} \\
	& = \frac{2\Omega_{n,\q}}{(i\Omega_m)^2 - \Omega_{n,\q}^2}
\end{align}
with $\Omega_m = 2\pi m/\beta$. To leading order the self energy is proportional to the diagram in Fig. \ref{fig:anharmbubble}, and it expressed as
\begin{align}\label{Polar_bubble_def}
	\Pi_n(\q=0,i\Omega_m) &= -\Tr\co{ \Lambda^{n,n_1,n_2}_{\textbf{k},\q=0} \, \mathcal{D}_{n_1}^{(0)}(\textbf{k} , i\omega_{m_1}) \pa{\Lambda^{n,n_1,n_2}_{\textbf{k},\q=0} }^\dagger \mathcal{D}_{n_2}^{(0)}(-\textbf{k} ,i\Omega_m-i\omega_{m_1}) } \\
	& = -\frac{1}{2}\beta \sum_{n_1,n_2}\sum_{\textbf{k}} \abs{ \Lambda^{n,n_1,n_2}_{\textbf{k},\q=0} }^2 \sum_{i\omega_{m_1}}\mathcal{D}_{n_1}^{(0)}(\textbf{k} , i\omega_{m_1}) \mathcal{D}_{n_2}^{(0)}(-\textbf{k} ,i\Omega_m-i\omega_{m_1})
\end{align}

\begin{figure}[h]
	\centering
	\includegraphics[width=0.35\linewidth]{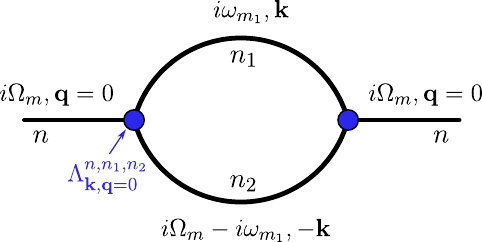}
	\caption{Phonon-phonon polarization bubble in the Matsubara representation. Here $n,n_1$ and $n_2$ are the moiré phonon branches and $\Lambda^{n,n_1,n_2}_{\textbf{k},\q=0}$ is the three-phonon vertex. As the mode $n$ must be Raman active, we take $\q = 0$. }
	\label{fig:anharmbubble}
\end{figure}

By performing the summation over the Matsubara frequencies $i\omega_{m_1}$, and taking the analytical continuation $i\Omega_m \rightarrow \Omega + i\delta$ the moiré phonon self-energy is written as
\begin{align}
	\Pi_n(\q=0,\Omega) = \frac{1}{2} \sum_{n_1,n_2}\sum_{\textbf{k}} \abs{ \Lambda^{n,n_1,n_2}_{\textbf{k},\q=0} }^2 &
	\left\{
	\co{n_B(\omega_1)+n_B(\omega_2)+1}\co{\frac{1}{\Omega -(\omega_1 + \omega_2) + i\delta} - \frac{1}{\Omega +(\omega_1 + \omega_2) +i\delta}} + 
	\right. \nonumber \\
	& \left.+\co{n_B(\omega_2)-n_B(\omega_1)}\co{\frac{1}{\Omega -(\omega_1 - \omega_2)+i\delta} - \frac{1}{\Omega +(\omega_1 - \omega_2)+i\delta}} \right\} , \nonumber
\end{align}
with $\omega_1 = \omega_{n_1,\textbf{k}}$ and  $\omega_2 = \omega_{n_2,-\textbf{k}}$. Finally, the dressed phonon propagator is written as
\begin{align}
	D_n(\textbf{q}=0,\Omega) = \frac{2\Omega_{n}}{\Omega^2 - \Omega_n^2 - 2\Omega_n\Pi_n(\Omega)}.
\end{align}
And the moiré phonon linewidth is going to be proportional to the imaginary part of the self-energy, given by:
\begin{align}
	\Im \Pi_n(\q=0,\Omega) = \frac{1}{2} \sum_{n_1,n_2}\sum_{\textbf{k}} \abs{ \Lambda^{n,n_1,n_2}_{\textbf{k},\q=0} }^2 &
	\left\{
	\co{n_B(\omega_1)+n_B(\omega_2)+1}\co{\delta\pa{\Omega -(\omega_1 + \omega_2)} - \delta\pa{\Omega +(\omega_1 + \omega_2)}} + 
	\right. \nonumber \\
	& \left.+\co{n_B(\omega_2)-n_B(\omega_1)}\co{\delta\pa{\Omega -(\omega_1 - \omega_2)} - \delta\pa{\Omega +(\omega_1 - \omega_2)}} \right\} \nonumber
\end{align}

A full numerical evaluation of $\Pi_n(\Omega)$ is computationally demanding, as it requires checking all kinematic constraints across multiple phonon branches and momenta. To make the problem more tractable, we adopt a series of reasonable approximations. First, we assume that the matrix elements $\Lambda^{n,n_1,n_2}_{\textbf{k},\q=0}$ vary weakly with both phonon branches and momentum. Under this assumption, the frequency dependence of the linewidth is approximated by the two-phonon density of states, $\gamma(\Omega) \sim \sum_{n_1,n_2,\textbf{k}} \delta(\Omega - \omega_1 - \omega_2)$. From this we see that, at low frequencies, the phase space available for decay is limited, resulting in sharper spectral features. As $\Omega$ increases, more scattering channels satisfy energy conservation, leading to increased broadening. Nevertheless, in the low-frequency regime relevant for the moiré phonon response, the linewidth remains nearly constant. Thus, within the low-temperature limit, it is justified to treat the broadening as frequency-independent.


\end{widetext}

\bibliographystyle{apsrev4-2} 
\bibliography{refs} 

\end{document}